# Beautiful secrets: using aesthetic images to authenticate users

Noam Tractinsky and Denis Klimov
Software and Information Systems Engineering
Ben-Gurion University of the Negev
Correspondence: noamt@bgu.ac.il

**Abstract**

We propose and evaluate an authentication scheme that improves usability and user experience issues in the authentication process due to its reliance on people's aesthetic tastes and preferences. The scheme uses aesthetic images to verify the identity of computer users. It relies on three major premises regarding visual aesthetics: (i) that an individual has different preferences for different aesthetic stimuli; (ii) that these preferences are relatively consistent; and (iii) that aesthetic tastes are subjective and, therefore, there are considerable individual differences in aesthetic preferences. Following a review of the scientific basis for these premises, we describe the concept of the aesthetic evaluation-based authentication (AEbA) method and illustrate an implementation of it. We address AEbA's advantages and disadvantages relative to other related methods and conclude that it is adequate for low-to-medium security domains. It cannot serve as a compulsory method because we suspect that a certain portion of the user population lacks the degree of aesthetic sensitivity required to use the system effectively. On the plus side, the method offers a positive experience. It alleviates the burden of memorizing passwords to a minimum, and relative to other usability-oriented schemes provides better security in terms of shoulder-surfing, phishing, and password space. Finally, we report on a pilot evaluation of the concept and its feasibility that supports the method's main tenets, provides insights about implementation challenges and suggestions for improvements.

## 1. Introduction

There is a consensus in the field of usable security that echoes Adams and Sasse's [1] call to consider the human link in the security chain. This recognition has led to a plethora of research on new authentication methods that will make authentication methods easier to use. Yet, in a comprehensive review of attempts to replace text passwords for general-purpose user authentication, Bonneau et al. [6] conclude: "Not only does no known scheme come close to providing all desired benefits: none even retains the full set of benefits that legacy passwords already provide. In particular, there is a wide range from schemes offering minor security benefits beyond legacy passwords, to those offering significant security benefits in return for being more costly to deploy or more difficult to use "(p. 554). Indeed, a recent laboratory study [59] found that participants still consider the password to be the most usable authentication method.

One way to address the inevitable trade-off between security and usability is to pursue security methods that are more customized and personalized [6, 60]. As noted by Garfinkel and Lipford, "security and privacy can be highly contextual, dependent on the person, the information, the problem, and use. This means there are likely few instances of one-size-fits-all solutions" [21, p. 108]. Against this backdrop, we propose a new authentication method, which is based on the evaluation of aesthetic images.

### 1.1. Proposed method

The aesthetic evaluation-based authentication (AEbA) method can improve the user experience (UX) in low- to medium-security authentication contexts. It relies on three central premises: (i) the things we like are sufficiently meaningful that they are easy to recognize and to be distinguished from things we do not like as much; (ii) aesthetic tastes are relatively consistent; and (iii) aesthetic tastes are subjective and therefore there are considerable individual differences in aesthetic preferences. Given the well-known trade-off between security and usability, we acknowledge that AEbA cannot serve as an all-encompassing authentication solution. Most importantly, it is not suitable for domains requiring very strong security. However, it has advantages over other methods for low-to-medium security domains. We elaborate on these advantages and limitations in Section 3.

How does AEbA work? Conceptually, an AEbA system includes a large database of images (e.g., photos), such as the images presented in Figure 1. In the setup phase, each user draws a random subset of those images and rates the degree to which he/she likes[1] them. It is important that all images are aesthetic (i.e., potentially likeable) to prevent low entropy distractors in the authentication process. In the authentication phase, the system retrieves and displays a *random* subset of images from the set of images that were rated by that user in the setup stage. Users are successfully authenticated if they select correctly from this subset the images that were highly rated by them in the setup phase. Depending on the configuration, the system may require that the user selects all and only the highly-rated images, or it can allow for some errors.

---

[1] We follow Skov [41], who equates "aesthetic appreciation solely with the assessment of liking, that is, how pleasurable or displeasurable a sensory experience is" (p. 23).

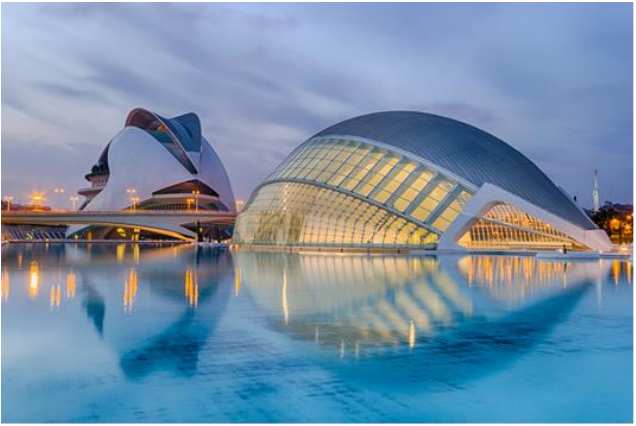
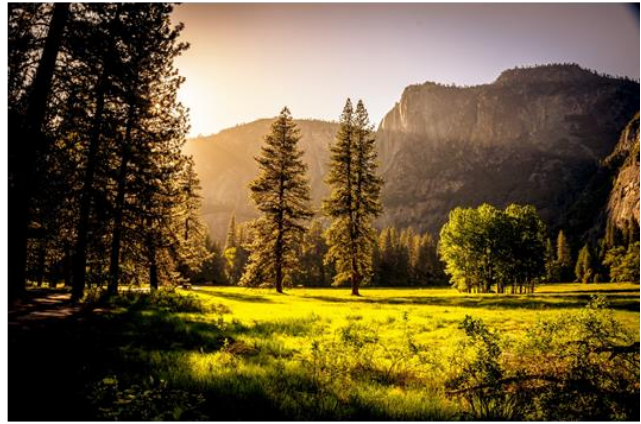
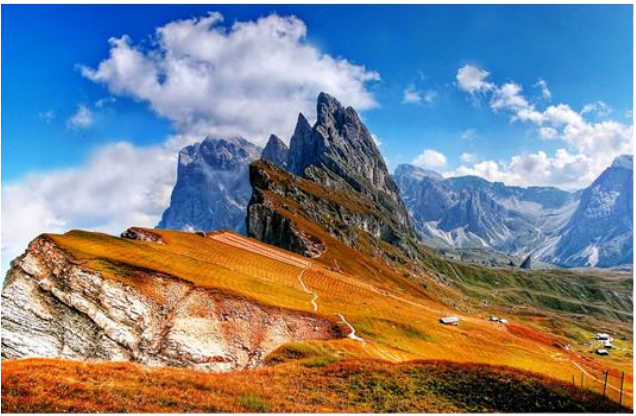
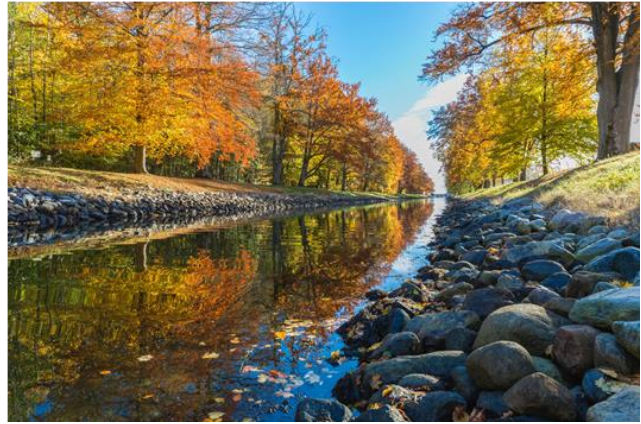

Figure 1: Examples of aesthetic images that could be used by an AEbA system (source: Pexels.com)

Superficially, the proposed method may be considered a member of a class of authentication methods that are based on users' knowledge, often termed *something you know (SYK)*, i.e., passwords and PINs. More accurately, though, the method relies on preferences rather than explicit knowledge. Similar to things we know, preferences reside in the user's mind. However, unlike SYK, preferences in their purest form are not necessarily based on explicit learning and memorization. Instead, they are part of who we are. Specifically, aesthetic tastes and preferences include affective and semantic contents; they are part of our identity [20]. They are compatible with our self-image [3], our values, and our positive and meaningful experiences. Because of that, they are relatively stable, and there is a high likelihood that they can be recalled and recognized easily and consistently, often implicitly and intuitively. In this sense, and because no two persons have exactly the same aesthetic taste and preferences, AEbA has some characteristics that resemble another authentication factor – *something you are (SYA)*.

AEbA focuses on improving the **usability** and the **UX** of the authentication process. Relative to arbitrary or random information that must be memorized and recalled (e.g., passwords), recognizing things we like is more efficient, effective, and pleasing, especially when it involves aesthetic images. In addition, while the proposed method is not aimed at top-security domains, it does have some security benefits relative to other methods with a similar password space. As all authentication schemes involve a trade-off between security and usability considerations [6, 21, 22], AEbA has advantages and disadvantages relative to other schemes. We elaborate on those in Section 3.2.

In the next section, we review the theoretical background for the AEbA method and relevant authentication methods to which AEbA can be compared. We then provide a fuller description of AEbA and report on a user study that assesses its feasibility. The final sections discuss the findings and summarize what we have learned thus far.

## 2. Background

In this section, we review the theoretical background that underlies the idea of AEbA, namely, using aesthetic evaluations and preferences as a basis for authentication. We focus on AEbA's main assumptions: the ubiquity of aesthetic evaluations, their subjectivity, and stability.

### 2.1. Aesthetic Evaluation

The history of visual aesthetics in human-computer interaction (HCI) is relatively short, yet its recognized importance to the field is growing [46]. Nevertheless, it is important to realize that even outside the field of HCI, past conceptions of visual aesthetics are being surmounted by new experimental evidence, especially from the nascent field of neuroaesthetics [10, 41]. Research in this field suggests that humans are routinely engaged in aesthetic evaluations of the environment [55], perhaps because beauty is linearly related to pleasure, a unique response curve among all other pleasures [9]. Research also suggests that the aesthetic experience emerges from the interplay of multiple neural systems with strong involvement of the reward circuit; it is highly subjective, and it involves not only the object but also semantics [11, 52, 12, 42]. While traditional aesthetic literature considered formal physical properties of the stimulus as the primary factors affecting aesthetic evaluation, the current view is that aesthetic evaluations are a result of actively comparing expectations, beliefs, prior experience, available information, and context [13].

Researchers agree that aesthetic emotions are experienced regarding any object that produces hedonic value [12, 32, 43]. There is also agreement that aesthetic experience is context-dependent, that is, a specific sensory stimulus acquires a specific hedonic value under specific conditions [25, 42]. Studies also stress that attention to the aesthetic experience and the feeling of pleasure as a companion of such experiences improve the memorability of aesthetic things [12].

As mentioned above, the use of aesthetic evaluations for authenticating a user relies on several premises regarding their characteristics: the superior memorability of aesthetic images; the consistency of those preferences within individuals; and the subjectivity of aesthetic evaluations, which leads to individual differences in preferences. We consider these issues below.

### 2.1.1 Memorability of images

The literature suggests that images are more memorable than words, aesthetic images are more memorable than ordinary images, and attentive evaluation of aesthetic images can further improve the memorability of images.

Humans have a massive storage capacity for images and for recognizing previously viewed images [45]. Relative to the capacity of verbal memory, we have a considerably greater ability to remember and to recognize images [40, 44, 33]. It was previously assumed that people mainly remember the gist of images, yet [8] demonstrated that people are adept at also remembering details of those images. Interestingly, studies have shown that the ability of experts (e.g., radiologists and cytologists) to recognize images from their domain of expertise dwarf in comparison to their (and ordinary people's) ability to remember general object and scene images. This suggests that memory for everyday images is considerably more capacious than for domain-specific images [18], presumably due to the already existing massive knowledge base of everyday images [7].

Research also suggests that it is more likely that people remember aesthetic images (whether those seen for the first time or those that have been seen before or that evoke past episodes). The reason is that aesthetic images induce emotional reactions, which enhance memory [30]. Emotions aid various stages of memory, including the process of encoding stimuli, storing, and retrieving them [35, 36, 50]. Although we do not expect the effects of images to be as strong as those of actual episodes, we still assume that the aesthetic experience they produce will have a positive effect on memory. This would especially be the case if images resemble positive autobiographical experiences (e.g., a great vacation in the Alps, a trip through the desert, or a visit to Paris). Indeed, some studies [cf. 2] have used judgments of emotion (pleasantness) to induce deep processing of images.

### 2.1.2 Stability of aesthetic evaluations

Relatively little research was done on the stability of aesthetic tastes. Pugach et al [38] found that stability at all age groups was rather low. However, the study employed a very small and homogeneous stimulus set. In a study of aesthetic-based preferences in abstract and real-world images, Vessel and Rubin [52] found that within-observer correlations (as a measure of consistency) had a mean of 0.67 for abstract images and 0.71 for real-world images. Similarly, Vessel et al. [53] found a test-retest reliability of around 0.7. Thus, the literature is yet indecisive about whether aesthetic preferences can serve as a reliable basis for authentication.

A possible reason for the lack of a clear answer to that question is that people are likely to differ in their ability to produce consistent aesthetic evaluations. Tractinsky et al. [47] report that the within-participants median correlation between two rounds of aesthetic evaluations of 50 images was 0.60, with a minimum of -.09 and a maximum of 0.90. However, it is important to note that evaluations in those studies were done under some form of time limit. For example, they compared evaluations in two rounds based on an exposure of 0.5 sec. and 10 sec. for the first and the second rounds, respectively. Thus, it is reasonable to conclude that there are individual differences in the consistency of aesthetic evaluations and that the benefits of AEbA will not translate to all users.

### 2.1.3 Individual differences in aesthetic evaluations

The idea that individual differences are at the root of aesthetic preferences is not new [31, 37]. Indeed, findings from recent studies support the notion that aesthetic evaluations differ between individuals. Vessel and Rubin [52] found that the agreement between individuals was higher for real-world images than for abstract images (mean pairwise correlations between different observers = 0.46 and 0.20, respectively, Experiment 1). However, under conditions that inhibited the use of semantic information about the images (i.e., observers relied mainly on visual features), the level of the agreement between individuals became considerably lower and essentially the same for abstract and real-world images [52, Experiment 2].

Similar findings are reported in [53]. There was greater agreement across individuals regarding natural classes of visual Images (e.g., landscape) compared to an agreement regarding artifacts, such as exterior architecture. Still, the study (Experiment 2) found that shared taste across individuals accounted for about one-third of the variance for natural landscapes and about one-fifth of the variance for exterior architecture. In addition, [53] found that only about one-fifth of landscape ratings and about 15% of exterior architecture ratings may be accounted for by low-level stimulus differences (i.e., luminance and color) indicating that much of the variance is produced subjectively rather than by objective or formal properties of the images. Iigaya et al. [24] claim the ability to predict people's tastes from a model of aesthetic preferences based on image features. However, their predictions of individual likings explain about 20% of the variance, leaving about 80% of the variance unexplained. Jacobsen and Beudt [26] concluded that of all different aesthetic domains, the highest variability in preferences is exhibited in the domain of visual aesthetics.

The neuroaesthetics literature raises a few possible explanations for differences in aesthetic evaluations. One such explanation involves the default-mode network (DMN), "a set of interconnected brain regions that are engaged by tasks that require inwardly directed attention or an assessment of self-relevance in the brain" [54, p. 19156]. This network is typically suppressed when a person engages with the external environment but was found in multiple studies to be highly engaged in aesthetic experiences across various domains [54]. Thus, substantial processing of the stimuli is done with relation to the observer's self-reference rather than with the stimulus's physical characteristics; therefore, preferences have a strong subjective component. This view is supported by research that found differences between brain regions that correspond to individual differences in evaluating facial attractiveness and preference for paintings, indicating the possibility that such differences originate at the neural level [10]. Another possible reason for individual differences is a considerable variation in aesthetic sensitivity between people [42] and in how people construct aesthetic preferences [13]. For example, people may prefer the mountainous sublime or sea sunsets, the high-rise building or the small cottage, classical or modern music and art, etc.

Obviously, individual differences in aesthetic evaluations make aesthetic preferences difficult to guess by adversaries. This is a desired characteristic of any authentication scheme and is a cornerstone of the AEbA method.

## 2.2 Related work on authentication

The AEbA scheme bears semblance to two families of authentication schemes. The basic idea that the key should be based on the user's liking is similar to the approach taken by preference-based authentication schemes. The reliance on images and the use of the challenge-response technique are attributes shared with the broad family of graphical-based authentication schemes. This section reviews both families of authentication schemes and discusses how AEbA is related to them.

### 2.2.1 Preference-based authentication

The idea of preference-based authentication (PBA) was introduced by Jakobsson et al. [28, 29], who recognized that personal preferences could reflect long-term, stable characteristics [4]. Things we like are ingrained in our minds, whether explicitly or potentially. By explicit, we refer to things that people are aware of liking, for example, places we like to be in or to visit, music, artwork, or foods that we grew to love. Potential liking refers to things we have not encountered yet, but which we will probably like given our tastes, values, and tendencies, or our explicit liking of similar things. For example, a new piece of music, work of art, or a dish may satisfy our tastes or capture our mind when we encounter them for the first time.

Originally, PBA was proposed in the context of password reset. In the setup phase, a large number of questions are asked regarding the user's personal preferences, such as "Do you like country music." Users respond on a 3-point scale, e.g., Really like, Don't care, Really dislike. In the authentication phase, the user is asked those questions again and is expected to answer them correctly, although few errors are allowed if they are small. Small errors are answers not at the opposite end of the answers provided in the setup phase. The users' responses are scored and are evaluated against a threshold score to determine whether the authentication succeeds. The findings in [28] suggest that very low false-negative and false-positive error rates can be achieved with 24 questions in the authentication phase.

In a subsequent study, [29] Jakobsson et al. suggested a method in which users had to select text-based items that they either liked or disliked from a larger set of items. Items not selected were considered by default to belong to the middle (Don't care) category. This allowed for a more efficient marking of items in the endpoints of the preference scale, given the assumption that users generally only have an opinion about a very small subset of issues. For setup, users selected eight items they liked and eight items they disliked. Item scores were based on the uncertainty of its answer for a random guess and were awarded or subtracted if the user's answers were correct or wrong, respectively.

Jakobsson and Siadati [27] replaced textual questions with images. In the setup phase, users are required to select a certain number of images they like or dislike from an initial set of images. In the provided example, the system shows 12 images simultaneously and requires that the user selects three images for "likes" and three for "dislikes." If the required number of likes and dislikes is not met, the screen is refreshed, and the process continues until the required number of liked and disliked images is achieved. Liking is based purely on semantic content (e.g., an image of a guitar or a hamburger), not on aesthetics. In the authentication phase, the user is shown a screen with images, randomly placed, and must select the same likes and dislikes that she selected in the setup phase. A limited number of small mistakes are allowed, based on the system's threshold. Obviously, greater tolerance for mistakes reduces the probability of false negatives and increases the likelihood of false positives. They report a registration process that takes 100 seconds and an authentication process that takes 40 seconds.

AEbA shares the PBA idea of relying on preference rather than on memorization. It also shares the idea manifested in [27] that images have unique properties that can make the authentication process more efficient. However, there are considerable differences between AEbA and the various PBS schemes.

1. AEbA is based on aesthetic evaluations, while PBA is indifferent to aesthetics even in its image-based version.

2. The potential number of PBA images is limited by the number of existing semantic objects or activities. In contrast, there is virtually no limit on the potential number of aesthetic images that can be included in an AEbA system. In fact, relatively subtle differences between apparently similar images could make a considerable difference in aesthetic perception, subsequently increasing uncertainty for potential attackers.

3. PBA requires only a coarse distinction between images, and therefore does not benefit from the addition of similar images. Thus, it contains a smaller number of items relative to AEbA. Consequently, the password space of a PBA application fits a limited range of authentication activities – basically password recovery and reset. AEbA, on the other hand, can fulfill this goal and, in addition, serve as a full authentication method for medium-security domains.

4. PBA is based on a 3-level differentiation on the likability continuum, while AEbA is based on a more fine-grained differentiation. This allows for more flexibility in determining possible keys in AEbA and consequently reduces the potential for false-positive error rates.

5. Setup time for AEbA is likely to be longer than the reported 100 seconds for PBA; on the other hand, authentication time will be shorter than the reported 40 seconds for PBA.

**2.2.2 Graphical-based authentication**

Relative to the conceptual similarity with PB authentication, the resemblance of AEbA to the broad family of graphical passwords is mostly technical. The idea of graphical-based authentication was introduced around 1999, "motivated by the promise of improved password memorability and thus usability, while at the same time improving strength against guessing attacks" [5, p.1]. A major claimed advantage of graphical passwords, which AEbA shares, is that they rely on the superiority of human memory for visual information compared to character-based memory. Consequently, proposals for various types of graphical passwords relied on the idea that they improve password *memorability*. Improved memorability, in turn, would motivate users to select more secure or less predictable passwords. Biddle et al. [5] classified three mechanisms of graphical password schemes – recall-based, cued recall-based, and recognition-based. Recall-based systems require that a person remembers information without cueing (e.g., a password or PIN). In graphical password systems, this is typically implemented by asking users to draw their password on a blank canvas or a grid [e.g., 16]. However, because recall is a difficult memory task, the cued-recall mechanism has been proposed to make password remembering easier for the user. Cued recall systems require that the user remember specific locations in an image. The system presents to the user an image that was introduced in the setup stage. That image provides a cue for the location of the correct answer. A prominent example of this type is PassPoints [57]. However, it is important to note that in graphical-based authentication, both recall and cued-recall require either recognizing a location or drawing/moving around locations when the image itself is not present on the screen. Consequently, these categories still face the challenge of memorizing arbitrary locations, which limits their usability.

The third mechanism relies on recognition, an easier memory task than recall [34]. Recognition-based graphical password (RBGP) systems provide information to the users and require them to decide if it matches the correct information. These systems usually use the challenge-response technique, in which users select an item from a displayed panel of images based on their knowledge of the secret. Still, RBGP systems usually "require that users *memorize* a portfolio of images during password creation and then must recognize their images from among decoys to log in" [5, p. 8, italics added).

Of the three graphical password types, recognition-based systems most resemble possible implementations of the AEbA method for three reasons: both methods use visual stimuli, both rely on recognition tasks, and both use the challenge-response technique to elicit the users' password. However, there is a crucial difference between the two. RBGPs rely on *arbitrary* stimuli that are usually devoid of affective content, must be *memorized,* and have little regard for the users' aesthetic tastes. AEbA, on the other hand, relies on the ability of users to select password keys (e.g., images) that they *like*. Hence, the need for memorization is diminished, and the probability of recognition is higher even without explicit or thorough memorization. From a security perspective, [5] suggest that recognition-based systems "are not suitable replacements for text passwords, as they have password spaces comparable in cardinality to only four or five-digit PINs" (p.8). However, AEbA's password space can be larger due to the advantages of using aesthetic images. It allows greater flexibility of images used, a more extensive database of images, and greater flexibility in the number of key and decoy images presented on one screen and the number of screens used in the authentication process. Nevertheless, given the similarities, we review below several notable RGBP systems.

Of the graphical-based authentication methods, the closest to AEbA is Déjà vu [17]. In Déjà vu, users select and *memorize* a subset of *random abstract* art images from a larger sample for their portfolio. To log in, they must recognize images belonging to their predefined portfolio from a set of decoy images. Importantly, [17] suggest that attractive images have to be included in the dataset to increase the likelihood of uniform distribution of selected images. The dataset in Déjà vu comprises random abstract art images only. They justify this choice because (i) abstract art can be easily remembered, and (ii) it will enhance security over a photograph-based portfolio that is easy to guess. However, the first argument stands in contrast to evidence from the neuroaesthetic literature, which indicates that images based on random abstract art are more difficult to remember relative to photographic images (e.g., [52, 7]). The second issue is easily mitigated in AEbA because, as noted above, individual preferences vary considerably when it comes to aesthetic images. Another difference between Déjà vu and AEbA is that Déjà vu uses a fixed number of 5 images as keys, whereas AEbA allows for a flexible number of key and decoy images.

Other recognition-based graphical passwords (RBGPs) are considerably less similar to AEbA. They all require memorization, mainly of arbitrary stimuli. Examples of such schemes are Passfaces [5], Story [15], the Convex Hull Click Scheme [58], and the Cognitive Authentication scheme [56].

A departure from this line suggests allowing users leeway to choose images in a non-arbitrary way (e.g., from their own photo archive or self-drawn doodles). Renaud [39] suggests that this improves the memorability of the password. Tullis and Tedesco [48] found that recognition accuracy for personal photos was higher than for stock photos. In a subsequent study [49], Tullis et al. found that recognition of personal photos remained high even after six years. However, using personal photos as passwords may come with the risk of personalized attacks if the users' style is known, as demonstrated by [48].

In general, recognition-based systems reduce the probability of phishing attacks, especially in schemes with variant responses[2], where only a portion of the user's secret is exposed on any login attempt. On the other hand, they are more susceptible to shoulder surfing, but this can be mitigated again by variant-response schemes [5], a method used in AEbA.

One of the problems identified in graphical passwords is that, like in regular passwords, users tend to select simple drawings, which decrease the variability of those passwords and make them more susceptible to adversarial attacks [21]. Another problem (and perhaps a reason for the previous one) is potential interference if such schemes are used in several systems, i.e., users' memory of a password for one system impairs their memory of a password for another system [5, 19]. These issues are less of a concern when implementing AEbA, which relies on innate preferences that do not depend on specific accounts or on memorizing images.

\*\*\*

To summarize, the literature supports the tenets on which the AEbA scheme is based and its potential superiority to comparable schemes. However, it appears that not all users can use AEbA because some do not have the required aesthetic sensitivity to produce consistent aesthetic evaluations.

### 3. Aesthetic evaluation-based authentication

In the previous sections, we presented the basic idea behind AEbA and provided a theoretical justification for why it could work. This section describes it in more detail and presents its advantages and limitations.

---

[2] In variant (non-static) response schemes, only a portion of the secret is exposed on any single login attempt and different items from the password portfolio are entered in different login instances.

## 3.1 A Conceptual Blueprint

Like most authentication methods, the AEbA concept is based on two phases: *setup* (enrollment/registration) and *authentication*. AEbA relies on a challenge-response technique that uses aesthetic images as stimuli. On the one hand, the combination of aesthetic images and the challenge-response technique supports easy recognition and a positive experience for the user. On the other hand, it increases the challenge to attackers.

In the enrollment phase, users' liking ratings of a large set of aesthetic images are elicited and stored in the authenticating system's database. In the authentication stage, the system draws from the rated stimuli two random subsets of the rated stimuli: highly liked stimuli (keys) and less liked stimuli (distractors/decoys). It displays the stimuli on the screen and asks the users to select the key stimuli from the displayed subsets.

To implement this image-based challenge-response system, consider a database containing several hundreds of aesthetic pictures, $AP$ (e.g., $200 \leq AP \leq 1{,}000$). All images must be aesthetic to prevent low entropy distractors in the authentication process. During the setup phase, users rate a subset $R$ of those pictures according to how much they like each picture (we consider a reasonable value of $R$ to be about 70, although a larger number of rated images is preferred). The ratings are stored on the authentication server. During the authentication phase, the system retrieves from the database a *random* subset of pictures from $R$, $D$ ($D > 2$), and displays them to the user. $D$ includes two *random* sunsets from $R$: a subset of pictures that were rated highly by the user, $D_{HR}$ ($1 \leq D_{HR} \leq D$), and a subset for which the specific user provided lower ratings, $D_{LR}$ ($D_{LR} = D - D_{HR}$). The random presentation of both the keys (highly rated pictures) and the decoys (not highly rated pictures) is an important feature of AEbA. It adheres to the recommendation in that "systems allowing some degree of user choice should encourage randomization of user-chosen sequences as well as individual items to avoid divide-and-conquer guessing attacks" [5, p. 34].

The strong version of the system verifies the identity of users if they select correctly *only and all* the highly-rated pictures presented by the system. Security can be boosted by using consecutive authentication screens ($S$) in a single authentication session. We consider a reasonable practical number of authentication screens in one session as ($3 \leq S \leq 5$).

One challenge facing the design of AEbA is the risk of false negatives. To minimize this risk, the system needs to verify that the recorded ratings of the presented key images are sufficiently distinct from the ratings of the distractor images.

In addition, relaxed versions that allow for some user errors (i.e., false negatives) are possible but are not discussed in this paper. We are less concerned about the need to manage false positives because *a priori* all images in the database are considered aesthetic, and hence attackers are unlikely to decipher a key based on this aspect. Thus, the attacker has to guess the user's aesthetic preferences, but as discussed above, this is unlikely due to individual differences in aesthetic preferences.

### 3.1.1 Usability and user experience

The usability benefits of AEbA as an authentication method are especially noticeable compared to various schemes using SYK methods. First, the combination of using aesthetic images and the challenge-response technique makes AEbA passwords more readily available for the user. Second, things that we like have positive emotional content, and hence the authentication process is more likely to create a positive authentication experience. This contrasts with various implementations of the SYK factor, which are character-based symbols with low semantic content, particularly those that constrain the use of characters in passwords or with other methods that have to be memorized out of context and hence are hard to remember and recall.

### 3.1.2 Security

The authentication literature agrees on the inevitable trade-off between usability and security [5, 21, 23]. Therefore, it is reasonable to expect that the improved UX of the AEbA will be accompanied by reduced security. AEbA implementation is flexible, though, and a typical implementation would provide theoretical security comparable to what we can find in a range between a randomly generated 4-digit PIN and an 8-character password [51]. For example, Table 1 shows the password strength in bits, given various values of the three main AEbA parameters.

Moreover, AEbA has several practical security benefits relative to SYK authentication types of comparable theoretical password space. First, aesthetic evaluations are more difficult to guess or extract by social engineering or shoulder surfing for two reasons: (1) they are fuzzier than traditional, true-false authentication factors, and (2) the response to the challenge is dynamic because the challenges (keys and distractors) are drawn at random from a larger pool of images and may change from one authentication session to another. Second, the mechanisms used for AEbA are not susceptible to phishing attacks as long as the authentication server itself has not been hijacked. Third, the correct answer to an AEbA challenge depends on the context (i.e., all the images that are presented on the authentication screen) because keys are based on liking images relative to other images. Consequently, knowing that someone likes an image does not necessarily help if we do not know how much the image is liked relative to other images in the displayed set. For ex-ample, we may know that a person likes mountainous scenes, but there may not be a photo of a mountain as one of the options in the set of photos presented to the user, or there may be several photos of mountains from which the user needs to select only a few. Or a mountainous scene may appear next to an option that is liked even better, say a sea sunset, about which the attacker may have no information. Thus, the truth-value of what the user likes is an unpredictable moving target for the attacker.

Table 1: Password strength in bits as a function of three main AEbA parameters: number of images displayed ($D$), number of images to select ($D_{HR}$), and number of screens shown in an authentication session ($S$), compared to the strength of memorized secret authenticators.

| $D$ | $D_{HR}$ | $S$ | Password space in bits | Memorized Secret authenticator |
|---|---|---|---|---|
| **6** | 1 | 5 | 12 | 4 digits PIN |
| **8** | 2 | 4 | 19 | 6-7 digits PIN |
| **10** | 2 | 4 | 22 | 6-7 digits PIN |
| **12** | 3 | 3 | 23 | 6-7 digits PIN |
| **12** | 5 | 5 | 49 | 8 characters (72-character set) |

**3.2 Synopsis: Advantages and Limitations of AEbA**

The main advantages and limitations of AEbA have been discussed in this and in previous sections. We summarize them here.

- AEbA does not require memorizing and remembering out-of-context words, images or spatial locations on the screen, let alone arbitrary characters.
- Relative to other challenge-response methods, AEbA is more resilient to phishing and shoulder surfing. This is because it emplos variant response, which makes the key fuzzy: correct and incorrect images are chosen randomly and changed from one authentication attempt to another. Thus, the key is not constant but rather determined by the relations of any image to the other images in the database.
- Relative to most SYK methods, AEbA is less susceptible to interference, because it does not depend on specific intersections of accounts and memorized secrets
- AEbA reduces the risk of users writing down passwords or communicating passwords to others.
- AEbA is scalable in terms of password space, depending on the parameters mentioned earlier (*AP,R,D,HR,LR,S*). Its effective range is comparable to schemes that effectively range between a 4-digit PIN and an 8-character password.
- The experience of using predominantly aesthetic images is likely to be more positive and pleasing than the use of arbitrary text or images.

Given the inherent imperfection of any authentication scheme, AEbA has its disadvantages too. These include:

- It is not suitable for domains requiring very strong security.
- It is not suitable for users who are not sensitive to visual aesthetics. However, it is essential to remember that no authentication scheme fits all users. The fact that various imperfect schemes (most notably, character-based passwords) are compulsorily employed on a large scale does not imply that they fit the strengths and limitations of the human cognitive system.
- The enrollment (setup) phase of AEbA may be relatively time-consuming (about 10 minutes). However, the authentication phase is expected to be reasonably short.

## 4. Evaluation and proof of concept

We conducted a study to evaluate the basic tenets of AEbA and the potential for its implementation. Specifically, the main goals of the study were to evaluate (1) whether people can distinguish between aesthetic images, (2) whether preferences of aesthetic images are consistent over time, and (3) the variability of users' aesthetic preferences of aesthetic images. The study was designed as a game that simulates the main attributes of AEbA, although it was not framed in terms of security/authentication. It received an expedited IRB approval and was conducted during a 3-week period in August 2021.

**4.1 Design and procedure**

To simulate AEbA, we developed a database of 318 images. Most images were downloaded from pexels.com – a free stock photo website -- and about 40 additional photos were taken from a personal photo archive. We classified the photos into eight categories (Universe, Nature, Mountains, Forest, Flowers, Cityscapes, Seaside, and Other). The study includes three main parts. The first two parts simulate AEbA's enrollment and authentication stages. The third part was designed to evaluate AEbA's resiliency to adversarial attacks.

*Stage 1: Image rating - Enrollment*. Following the approval of a consent form and a short registration procedure, the system displayed a screen with 16 thumbnailed photos selected from all the photo categories to acquaint the users with the type of images they will encounter during the game. All images were presented in a horizontal (landscape) layout. Next, the system randomly chose 72 images from the database and presented them one by one. The users rated each photo according to how much they liked it on a 1-10 scale (1 = don't like, 10 = like a lot). Figure 2 shows a screenshot of the rating process. Users were not allowed to return to previously rated photos; thus, they could not change their scores or compare different images or scores of images. Upon completion of the ratings of the 72 photos, the participants could choose between rating additional photos or moving to the next stage. If the participant chose the first option, the rating procedure would continue with randomly drawn images until the participant ended it. The participants could at any time return to the rating stage to continue rating additional images, but they could not see again images that they had rated previously.

*Stage 2: Authentication game.* Participants were eligible to enter the second stage (which we termed "The Game") after completing the ratings of at least 72 images. They could enter the game following the rating session or directly after logging into the system at a different time. The game included four screens, each screen holding eight thumbnailed images, of which two were among the top 20% rated images by that user and six were among the bottom 60% rated images. Users were instructed to select their two most preferred images out of the eight presented images on each screen. Selections were compared to the ratings in Stage 1, and for each correct selection, the participant gained one point. Thus, the participants could have earned up to eight points in each game session. Figure 3 shows a screenshot of the first screen with feedback given to the user (two correct selections). On the final screen, the system also informed the users about their total score in the trial.

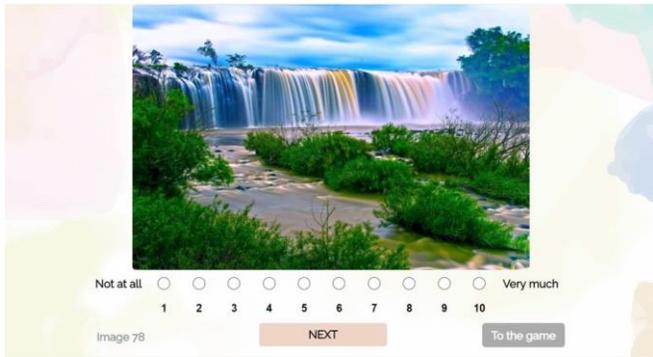

Figure 2. An image rating screen. Users were instructed to rate each image according to how much they liked it.

To test aesthetic consistency rigorously, we made two design decisions. First, we did not want the participants to learn about the accuracy of their selections, because that would taint the consistency tests. Thus, if users selected the wrong images, they did not receive specific feedback about which selection was wrong or which image they should have chosen instead. Second, we wanted the participants to play the game for several weeks. However, we restricted them to playing no more than one session a day to reduce the effects of memorization and repeated exposure on the results while maintaining their interest and involvement. To further motivate the participants, they were allowed to access the game's leaderboard at any time.

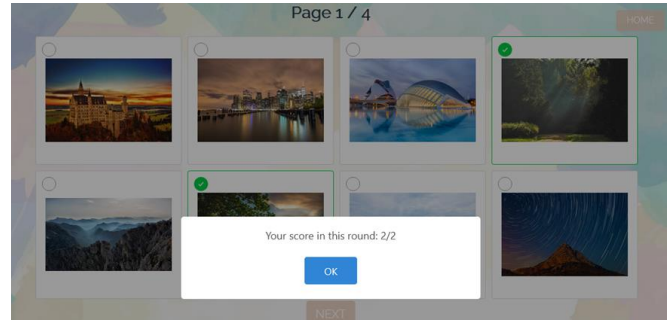

Figure 3. User-selected images are marked in green. Upon confirming the selections, the user receives feedback about the score in the current screen (first screen out of 4, in this example). If there are incorrect selections, the system provides no feedback on which of the selections is wrong or which should have been chosen instead.

*Stage 3: Adversarial game.* At a certain point after the beginning of the experiment, we informed the participants that they could play another game (we used the term "Advanced Game"). The game involved guessing other players' aesthetic preferences. The purpose of the second game was to serve as a measure to assess the variability of aesthetic preferences among our sample and to evaluate empirically the probability that adversaries can successfully guess the passwords of other (unfamiliar) users. From the participant's perspective, the adversarial game was conducted the same as *The Game* (Figure 3). The only exception was that in each session the presented images were taken randomly from a pool of sessions played previously by other players. There was no limit on the frequency of playing this game. This game had a separate leaderboard, which the participants could access at any time.

### 4.2 Sample

We recruited a convenience sample[3] of 44 participants for the study. We used the data from all participants to analyze the image ratings but excluded from the analyses of the game 11 participants who completed less than two sessions, leaving a total of 33 eligible players (21 females and 12 males).

---

[3] Convenience sampling is common in exploratory evaluations of the types used here. A major drawback of convenience sampling -- acquiescence bias – is irrelevant here because we are interested in performance measures that can not exceed the human's ability regardless of the participants' motivation and effort.

## 4.3 Results

### 4.3.1 Ratings

Overall, 274 images were rated by the study participants, for a total of 3722 ratings. The average rating was 6.07, the median was 6, and the most frequent values were 7 and 8. Figure 4 shows the rating distribution, which is skewed to the right in line with the aesthetic nature of the images.

For further analysis, we excluded eight images that were rated by less than 5 participants. The average rating of the remaining 266 images was 6.12 (SD=1.23). The least liked image had an average rating of 2.32, and the most-liked image's rating was 8.63. Figure 5 presents the average rating of each of the 266 images, together with the minimum and maximum rating values for each image. The difference between the minimum and maximum values averaged over all images was 7.0. As can be seen, the average ratings of almost 80% of the images lie between 5 and 8. Most importantly, within this range, 197 images received a maximal rating of at least 9, and 161 images received a minimal rating of 3 or lower. These results indicate a large variance in the participants' tastes and preferences.

In terms of individual ratings, 31 out of the 44 raters used the full 10-point scale (1-10) to denote their liking of individual images. Eight raters used a 9-point scale (that is, they did not include one of the end-points of the scale). The other five participants used a shorter liking scale but no less than a 6-point scale. These findings indicate that most users were able to assign highly differential liking scores to the images presented to them.

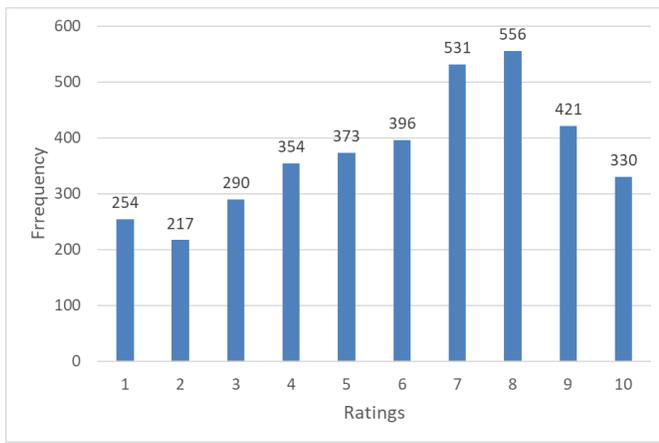

Figure 4. Distribution of all 3722 image ratings (1 = don't like, 10 = like a lot) in the study's first stage.

### 4.3.2 The Game

After excluding 11 participants who played the game only once, we analyzed the data of the 33 participants who completed the game at least twice. In total, we recorded 264 playing sessions. On average, those participants played eight sessions of the game, yet the number of games played varied considerably between 2 and 20, with a median of 6. The average success rate of the participants was 6.06 out of 8 (76% success rate). Figure 6 shows a box plot chart of the distribution of the game's scores for the 33 participants who have completed the game at least twice. The linear correlation between the order of the game and its score was low and negative (r=-0.17), indicating a slight decline in performance over time. This decline was less pronounced among the top-half (seventeen participants) performers (r=-.09) and completely disappeared among the top third (11 participants) performers (r=.01).

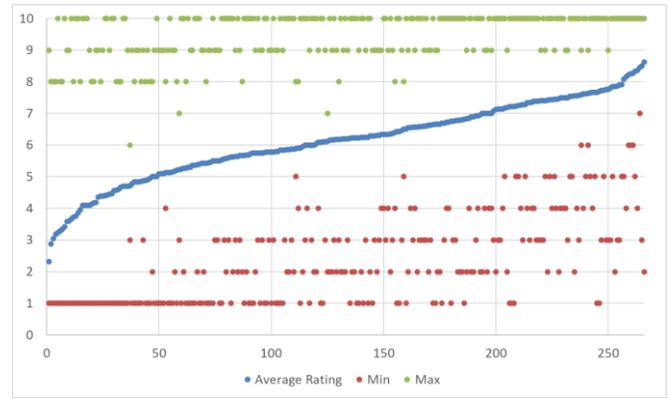

Figure 5. Ratings of study images (minimum of 5 ratings per image). Images are sorted in ascending order of average rating. Each image is represented by a blue dot (average rating), a red dot (the image's minimum rating), and a green dot (the image's maximum rating).

### 4.3.3 The adversarial game

We opened the adversarial game for participation ten days after launching The Game. Consequently, the participants had less time to play that game. However, unlike "*The Game*," there were no restrictions on the frequency with which they could play the adversarial game. Overall, 209 games were played, out of which 190 results were recorded (19 games did not record results due to technical problems). Figure 7 shows the distribution of correct guesses in those games. The average number of correct selections in this game was 2.88 for a 36% success rate. In only seven games (3.7%), participants guessed at least 75% or more of the correct images.

### 4.3.4 False-negative and false-positive errors

We calculated the probabilities of making wrong decisions due to identifying legitimate users as imposters (false-negative errors) and imposters as legitimate users (false-positive errors). False positives (in orange) and negatives (in blue) for the entire data set are plotted for each possible outcome of the two games, as indicated on the abscissa in Figure 8. The value of false positives was calculated as the proportion of sessions reaching that outcome in the adversarial game. The value of false negatives is (1- proportion of sessions reaching at least that outcome in The Game). The overall trend of the two curves in Figure 8 is in the expected direction. However, a practical authentication application would need to reduce the false negatives rate considerably. Still, if we only consider the top-half performers in the evaluation (17 participants, yellow curve), then the false-negative rate decreases considerably. Considering the top 1/3 (gray curve) the rate decreases even further.

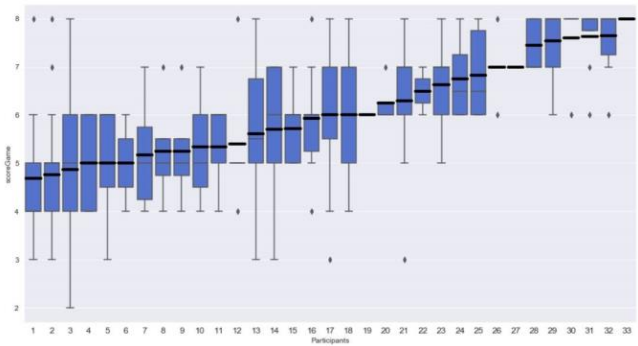

Figure 6. A box plot chart of the distribution of game scores of 33 participants who completed at least two games. Mean scores are denoted by a bold black horizontal line. For each participant, the plot displays the following values of game scores: median, 1$^{st}$ and 3$^{rd}$ quartiles, minimum, maximum, and outliers.

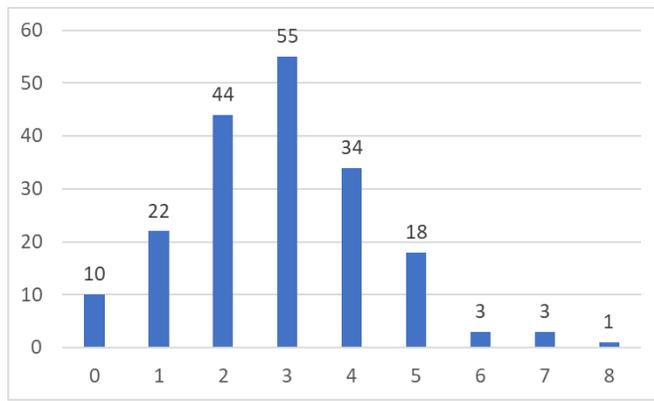

Figure 7. Distribution of correct responses in 190 games in which participants guessed the most liked images of another player.

#### 4.3.5 Possible issues with the study's procedure

We looked into possible issues that have hampered performance in the game (as indicated in Figure 8). Recall that in the first stage of the game, we did not allow users to change their ratings or see previously rated images or compare various images and ratings. We believe that the serial nature of the rating procedure could have caused inaccurate relative ratings. The reasoning (supported by participants' feedback – see Section 4.3.6) is that users may find out during the rating process that images that received a particular grade should have been graded lower or higher relative to new images that now appeared for rating. However, our application did not allow them to revise their ratings. Consequently, inaccurate ratings in Stage 1 had a deleterious effect on users' ability to distinguish between keys and decoys in the game (Stage 2). In the next section, we consider how to modify the rating procedure to prevent this effect.

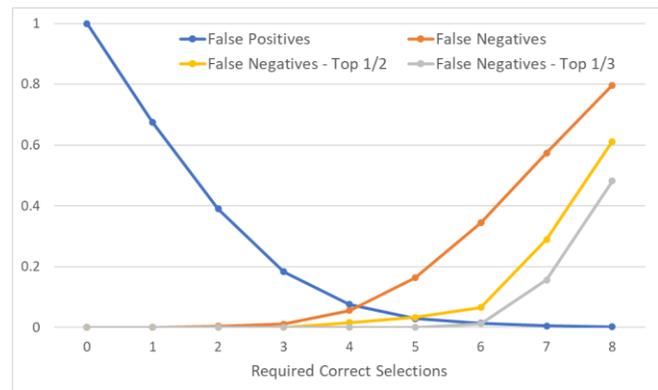

Figure 8. False positives and false negatives for each number of correct selections (X-axis) in games simulating an authentication application. The blue curve indicates false-positive rates for the entire sample (based on the adversarial game); the orange curve indicates false-negative rates for the entire sample (based on "the game"); the yellow and gray curves indicate false-negative rates for the top-half and top-third performers in "The Game," respectively. The rate of false positives indicates the proportion of users who could have been authenticated by mistake if that were the threshold for authentication. The rate of false negatives at any point indicates how many legitimate users would not be authenticated if that point served as the threshold for authentication.

Another issue is the rating distributions of the participants. To illustrate this issue, Figure 9 juxtaposes the rating distributions of the three most successful players (blue charts) and three of the least successful players (orange charts). The charts indicate that successful players had balanced rating distributions, especially at the high end of the rating scale. Such distributions work best with the mechanism AEbA uses to present keys and decoys in the game. Conversely, the rating distributions of unsuccessful players were unbalanced in various ways. Indeed, given the game's image selection mechanism in the authentication phase, which requires roughly a ratio of 1:1:3 between key, buffer, and de-coy images, respectively, such unbalanced distributions harm performance.

#### 4.3.6 Participants' comments

We approached some of the participants for their reactions to the games. While most of them considered the experiment an inconsequential game, they indicated that they could have improved their performance if allowed to revise their initial ratings. Some of them suggested that their ratings became consistent only after viewing about 20 to 30 images. That is, it took about this number of images for their ratings to become calibrated with their actual feelings about the images. We had hoped that showing a screen of 16 thumbnailed images at the beginning of the process would serve this calibration purpose, but apparently, only watching several images was not as informative as actually rating images.

Participants diverged in terms of their enthusiasm for the game. Some found it unattractive, while others enjoyed it very much. Belonging to this second group were members of a 3-generational family, ranging from teenagers to people in their seventies, who eagerly played both games and competed against each other.

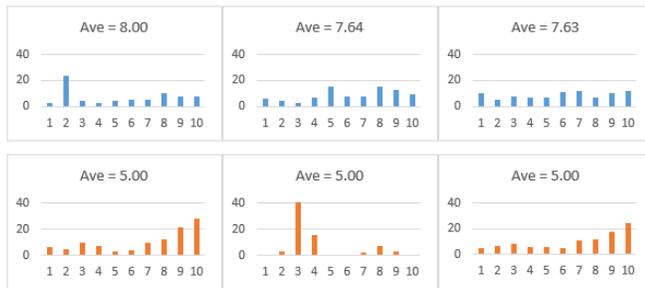

Figure 9. Distributions of image ratings by top performers (top three panels, in blue) appear more balanced than those of low performers (bottom three panels, in orange). Chart titles indicate the users' average performance in the game.

## 5. Discussion and future work

The pilot study results provide strong evidence in support of two of the basic tenets on which AEbA rests, but the evidence was weaker for the third one. Our first tenet was that people have differential preferences for aesthetic images. The results show that participants used a wide distribution of grades to rate the various images that they saw. Even when facing a set comprised of predominantly aesthetic images, most users could still distinguish between images they liked more and images they liked less, assigning liking ratings across the entire rating scale (e.g., Figures 4 and 5).

The AEbA scheme also assumes diverse interpersonal aesthetic tastes; given a large pool of aesthetic images, different people will exhibit different preferences. This assumption was supported by two different findings: the broad distribution of ratings of most images (Figure 5) and the difficulty of people to guess other people's preferences (Figure 7).

Finally, AEbA depends on users' ability to maintain consistent aesthetic preferences over time. The pilot study showed that about one-third of the participants maintained consistent preferences, with only one out of 33 users performing perfectly. For the other participants, consistency declined slightly over time. In addition, some users performed relatively poorly in the game. This finding was expected given the individual differences mentioned in Section 2.1.2. Indeed, the results of the top-half performers in the game were much better than the bottom half.

As mentioned earlier, several aspects of the game's procedure hampered the users' ability to play a perfectly consistent game. During the rating stage, the procedure did not allow users to return to previously rated images, change their scores, or compare different images and their scores. These restrictions were also notable in comments made by the users in exit interviews.

In addition, we found that users were more consistent if they could roughly maintain a balanced distribution of image ratings (Fig. 9). We are not yet clear about the source of differences in rating distributions between users. It could be that specific users have a harder- or easier time dealing with the rating instructions. It could also be that unbalanced distributions stem from users' lack of aesthetic sensitivity or from attributes of the image pool from which images are selected for presentation.

The next version of the system will be designed to accommodate these issues. The main change will be a modification to the rating system that will allow for parallel viewing and rating of the images at all times during the rating stage. Such a design will enable the revision of one's ratings, a feature that did not exist in the current version. We will also use instructions and visual cues to nudge users into more uniform rating distributions. We expect these modifications to improve the performance of both high- and low-performers.

Reflecting on the entire project, we also realized the importance of pretesting the image bank before deploying a game (or a service) to ensure that it does not contain images with low rating dispersion or images which are rated (on average) on the extreme ends of the liking scale. Eliminating such images should make guessing other people's preferences even harder and hence further reduce the false positives rate.

In the longer run, we consider several research venues to further develop and enrich the AEbA concept. One such direction is to examine how variations in the different pa-rameters of AEbA and changes to the image-selection algorithm affect the method's effectiveness.

The resemblance of some AEbA characteristics and the *something you are* authentication type may open the door to various additional approaches to authentication. In general, the combination of authentication by preference (or liking) and aesthetic tastes can be explored further, not only in terms of visual stimuli but also in terms of involving the aesthetics of other senses. This suggestion is based on findings from the neuroaesthetics literature [42,43] that pleasure triggered by senses other than the visual create similar brain activity patterns in the reward system.

It would also be interesting to study the potential of using machine learning techniques to predict individual users' evaluations and to generate keys and decoys that the user has not previously rated. Doing so could increase the password space by increasing individual users' image pools and their variability.

Finally, AEbA may have additional advantages for specific populations. The advantages associated with the greater ease and accuracy associated with two AEbA features – the use of aesthetic images instead of text and recognition instead of recall. For example, studies suggest that the advantage of recognition over recall is even more pronounced with aging populations [14].

## 6. Conclusion

Security is of utmost importance for high-risk domains, such as online banking or national security. Graphical passwords are not suitable for such domains because their password space is not large enough to protect the integrity of the authentication process. However, in lower-risk domains, more memorable schemes can improve user experience and eliminate prevalent risky behaviors such as writing down passwords or using the same password for both high-risk and low-risk accounts [5]. More memorable passwords are especially important for infrequently used accounts because memory decays over time. This paper proposed AEbA, an authentication scheme that relies on users' aesthetic evaluations of images. We suggested that for those low to medium-risk domains, AEbA provides considerable advantages (see summary in Section 3.2). Most notable, it improves the usability and user experience of the authentication process by alleviating the need to memorize and recall passwords.

We conducted a proof-of-concept study to assess the feasibility of using aesthetic evaluations as a basis for authentication. The results were encouraging and pretty much in line with our expectations. We noted several possible modifications to the application which may help reduce false negatives and increase the potential user base of an AEbA service. Future research should further explore improvements and extensions to the AEbA concept.

## Acknowledgements

We thank Shir Ben Dor, Yarden Levy, Liron Oskar and Taly Schvartz for developing the software for the PoC study.

## A   Image Set

To illustrate the type of images used in the evaluation study, we provide three samples from that dataset. Based on our study we tentatively consider a threshold of 4.00 to be an appropriate lower bound and 8.00 as an upper bound for inclusion in the AEbA dataset. The first sample includes images with low average ratings of just *above* the lower bound threshold of 4.00. The second group of images received an average rating around the study's mean (6.07). The third group received average ratings just *below* the upper bound threshold of 8.00. As can be seen, almost all images received extreme ratings from individual participants. These findings support the notion of high variability in individual preferences.

**Group 1: Eleven lowest rated images with average ratings > 4.00**

Legend for values below each image: Average, Min (lowest rating of that image), Max (highest rating of that image), Number of ratings.

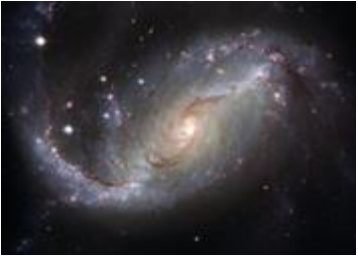
4.09, 1, 10, 11

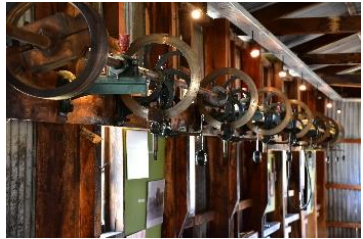
4.10, 1, 10, 21

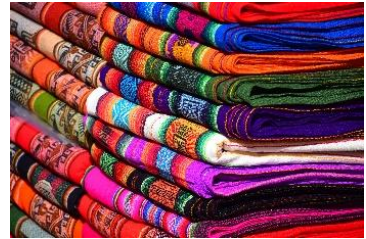
4.10, 1, 9, 10

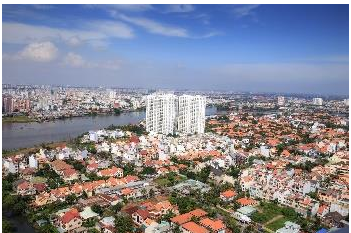
4.13, 1, 8, 8

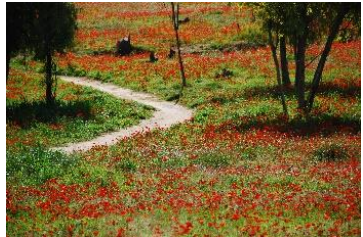
4.17, 1, 8, 6

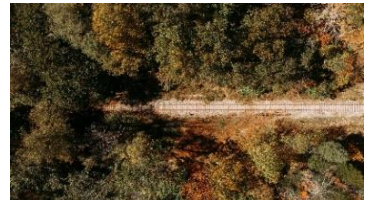
4.18, 1, 9, 11

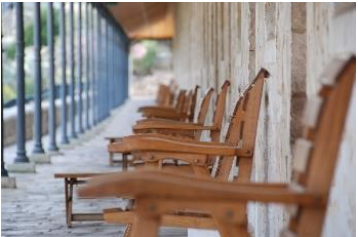
4.36, 1, 9, 22

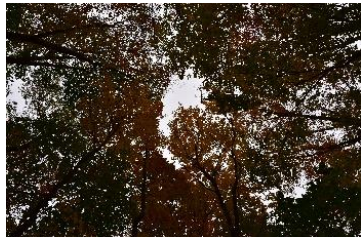
4.38, 1, 8, 13

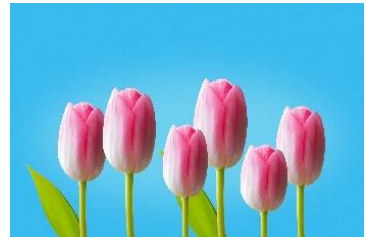
4.39, 1, 9, 18

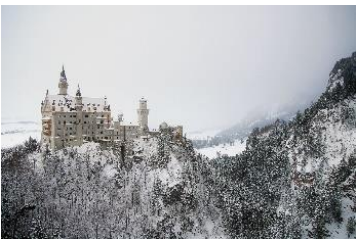
4.40, 1, 10, 10

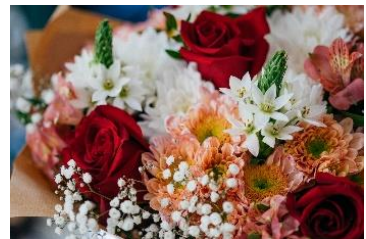
4.41, 1, 10, 17

**Group 2: Ten average images: images with ratings around the sample's mean of 6.07**
Legend: Average, Min, Max, Number of ratings.

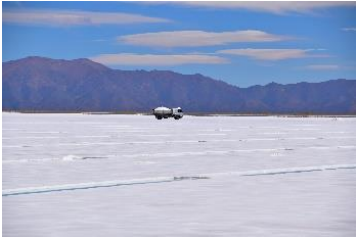
6.00, 3, 10, 16

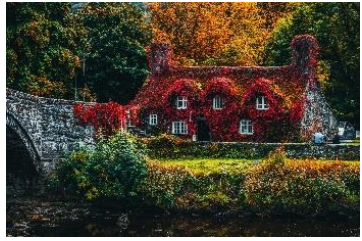
6.00, 4, 10, 9

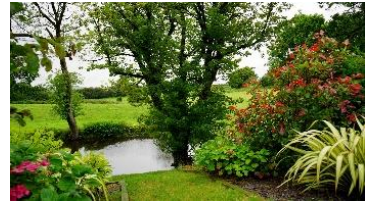
6.00, 2, 10, 9

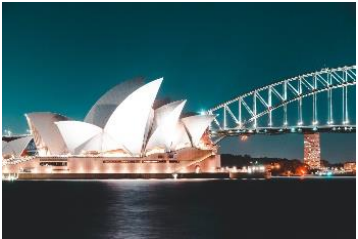
6.04, 2, 10, 23

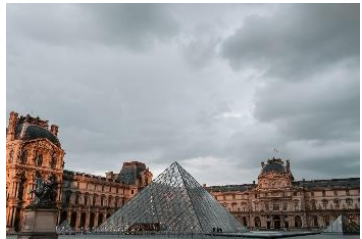
6.08, 4, 9, 12

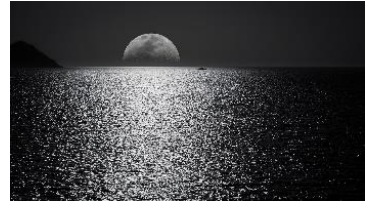
6.08, 1, 10, 12

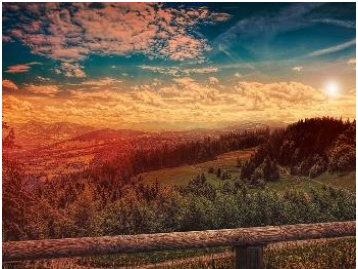
6.11, 1, 10, 19

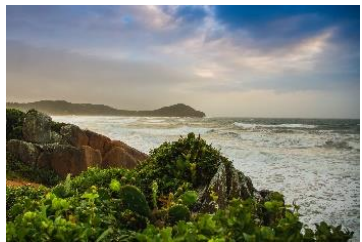
6.11, 3, 9, 9

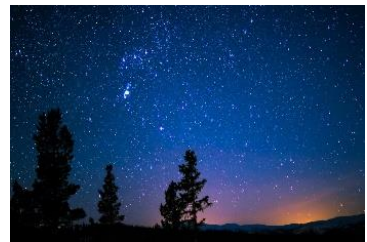
6.13, 2, 7, 8

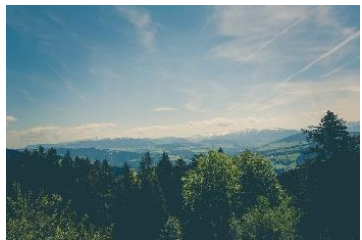
6.14, 2, 10, 14

**Group 3: Top 11 images with average score < 8.00**
Legend: Average, Min, Max, Number of ratings.

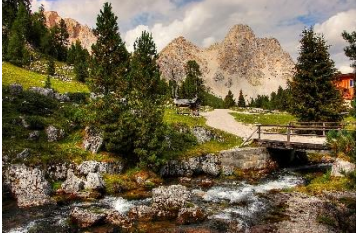 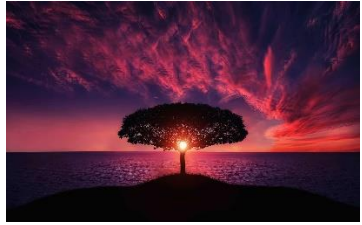 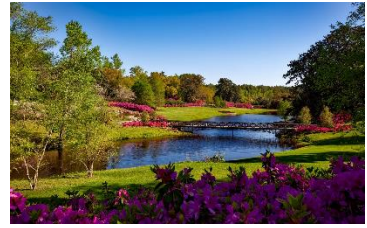

7.67, 5, 10, 9      7.67, 1, 10, 12      7.73, 3, 10, 15

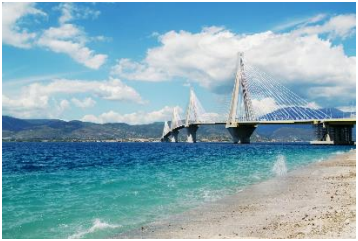 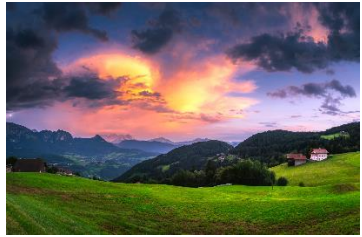 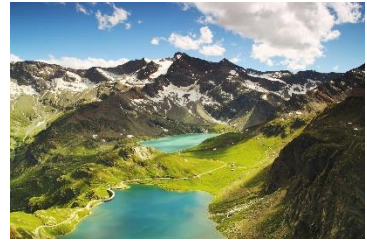

7.73, 5, 10, 15      7.75, 3, 10, 8      7.75, 3, 9, 12

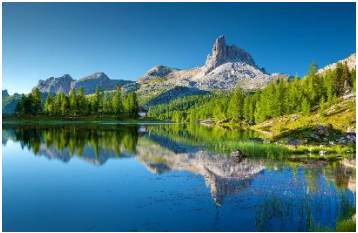 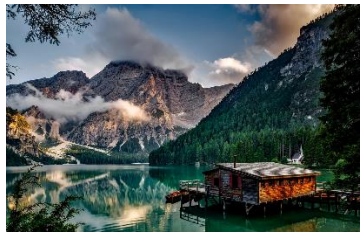 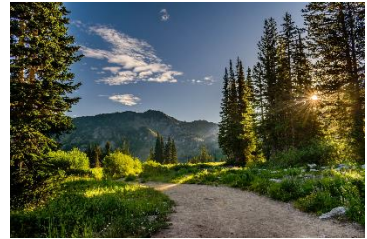

7.80, 3, 10, 15      7.86, 5, 10, 14      7.86, 2, 10, 14

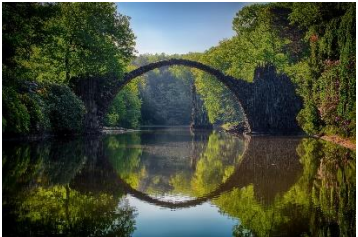 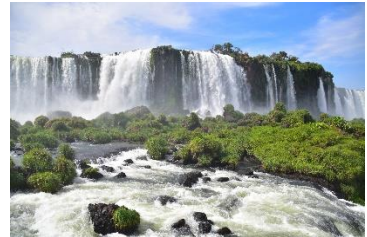

7.87, 3, 10, 15      7.91, 5, 10, 11

**For comparison, below are the three images that received the highest average ratings. Nevertheless, two of these images still received some very low ratings.**

Legend: Average, Min, Max, Number of ratings.

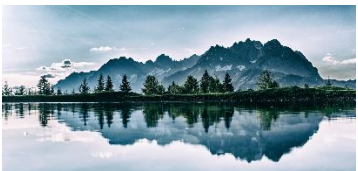 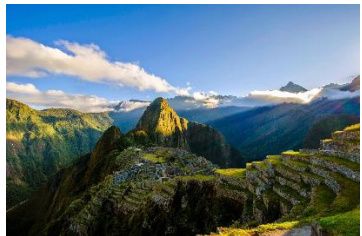 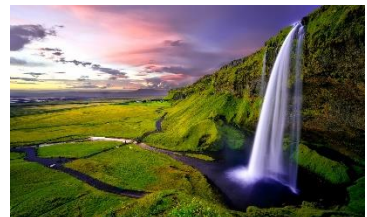

8.45, 7, 10, 11      8.50, 3, 10, 12      8.63, 2, 10, 8

# B  Game Instructions

Screens 1 and 2 are mandatory for first-time users of the system. First-time users can continue to screens 3-9 immediately after reading the instructions on Screen 2. If they have not finished rating at least 72 photos, they must return to Screen 3 on their next session. If they rated at least 72 photos, they could return in their next visit directly to Screens 3 or 8.

**Screen 1: Entry and Registration**

Welcome to the photo preference game.

The game is being conducted as part of a research project at the Anonymous Department.

The game is divided into two stages:

Stage 1. You will watch 72 photos and rate how much you like each one. This stage will last about 10 seconds. Your ratings must accurately represent your preferences because your success in the game depends on your ability to recognize which photos you liked more and which ones you liked less.

Stage 2. After that, we invite you to play the game. Whenever you enter the application, you will be shown four screens. Each screen will include eight photos from those you have rated in Stage 1. The challenge is to identify and select in each screen the two photos for which you provided the higher rating in Stage 2. Each correct selection will award you one point.

The goal of the game is to accumulate more points. Please do your best to make the game competitive.

We invite you to play once a day during the game's availability period. You cannot play the game more than once a day.

To help us administer the game and to communicate with you during the game period, we ask you to provide your email address to which we would send messages (for example, a reminder that you did not play the game during the last few days). There is no need for any identification information beyond this. You can also choose a nickname that will be shown on the leaderboard with your points total.

We guarantee:

1. To update you about your achievement when the game ends.

2. That all of the game's data will be saved on a secure computer and will only be used for statistical analyses. None of your personal data will be published.

[ ]  I agree to participate in the photo preference game according to the above conditions.

**Screen 2**

To begin the game, you should view 72 photos and rate them according to how much you like them.

Here you can view a sample of thumbnail photos that gives an impression of the type of the game's photos. Once you are done viewing the photo sample, press NEXT.

You will then be presented with a larger photo. At the bottom of the screen, please indicate how much you like the photo. To continue to the next photo, press NEXT. This is how your rating of the 72 photos will be recorded.

Once you have finished rating all required photos, you will be able to choose between the rating of additional photos and moving on to the game.

**Screen 3**

Now, please rate how much you like each photo. We recommend that you use the entire scale (from high scores for photos you like very much) to low scores for photos you least like.

**Screen 4: Rating screen (see Fig. 1 in the paper).**

**Screen 5: Pause and informational screen**

Excellent. You have rated 24/48 photos out of the required 72 photos. You can continue rating photos by pressing the Next button.

**Screen 6: Finish mandatory rating screen**

Great! You have now rated 72 photos, which is the required minimum for participating in the game. The more photos we have with your ratings, the better your chances of succeeding in the game. If you would like to rate additional photos, please press the Continue Rating button. You can stop rating at any stage by pressing the Finish Rating button. To start playing the game, press the To the game button.

**Screen 7: Game instructions**

The game, at last! We will now show you four screens. On each screen, there will be eight thumbnails of the photos that you have rated in the previous stage. Of the eight photos, select the two that you most like. The goal of this game is to select the photos that received the highest ratings in the previous stage. For each correct selection, you will be awarded one point. To unselect a selected photo, click on it again.

**Screen 8: The Game (see Fig. 2 in the paper).**

**Screen 9: End of the game session**

Game over! You earned X points today. Your overall score is Y points. See you tomorrow!

**For returning players:**

**Screen 10: Returning players**

Welcome *<nickname>*,

You have last played on *dd/mm/yyyy*.

You have won <y> points thus far.

To start the game, click *To the game*. [This option advances the user to Screen 8]

To rate additional photos, click *Rate photos*. [This option advances the user to Screen 3]